\documentclass[9pt,twocolumn,twoside]{opticajnl}
\journal{opticajournal} 

\setboolean{shortarticle}{true}

\usepackage{braket}
\usepackage{physics}
\usepackage{comment}
\usepackage{siunitx}
\usepackage{xcolor}
\usepackage{lineno}
\usepackage{ulem}

\newcommand{\pll}{{\mkern3mu\vphantom{\perp}\vrule depth 0pt\mkern2mu\vrule depth 0pt\mkern3mu}} 
\newcommand{\vect}[1]{\ensuremath{\mathbf{#1}}} 

\linenumbers 

\title{Computational refocusing in phase-resolved confocal microscopy}

\author[1,2,3]{Martin Schnell}
\author[4]{Melanie King}
\author[5]{Sam Buercklin}
\author[1,6]{Paulo Sarriugarte}
\author[1,3,7]{Rainer Hillenbrand}
\author[4*]{P. Scott Carney}

\affil[1]{CIC nanoGUNE BRTA, 20018 Donostia--San Sebastian, Spain}
\affil[2]{Beckman Institute for Advanced Science and Technology, University of Illinois Urbana-Champaign, IL 61801, USA}
\affil[3]{IKERBASQUE, Basque Foundation for Science, 48013 Bilbao, Spain}
\affil[4]{The Institute of Optics, University of Rochester, 480 Intercampus Drive, Rochester, NY 14620, USA}
\affil[5]{Department of Electrical and Computer Engineering, University of Illinois Urbana-Champaign, IL 61801, USA}
\affil[6]{Department of Applied Physics, School of Engineering Bilbao, University of the Basque Country (UPV/EHU), Torres Quevedo Plaza 1, Bilbao 48013, Spain}
\affil[7]{Department of Electricity and Electronics, UPV/EHU, 20018 Donostia-San Sebastian, Spain}

\affil[*]{Corresponding author: scott.carney@rochester.edu}

\begin{abstract}
We demonstrate numerical refocusing in coherent confocal laser scanning microscopy based on synthetic optical holography. This physics-based approach implements a computational propagation on the complex signal recovered in synthetic holography consistent with the wave physics and the parameters of the microscope. An experimental demonstration is shown to restore an in-focus image of a test object from data acquired at several focal plane off-sets. Numerical refocusing can provide focused views on samples with large height variation with a potential application in confocal optical surface profiling.
\end{abstract}

\setboolean{displaycopyright}{false} 

\begin{document}
\nolinenumbers
\maketitle

\section{Introduction}
Quantitative phase imaging plays an important role in applications such as non-contact surface profiling of micromechanical structures \cite{RN66,RN194} and imaging of transparent biological objects \cite{RN9,RN10}. Knowledge of both amplitude and phase of the scattered field also allows for computational processing of the recorded images based in wave physics, thus enabling numerical refocusing on different object planes and the correction of aberrations after image acquisition \cite{RN14,RN15}. Driven by these advantages, digital holography has been an important area of research in wide-field microscopy and led to the development of commercial products.

Confocal microscopy offers superior contrast and depth discrimination as compared to wide-field techniques and finds wide application in fluorescence and Raman imaging. Interferometric methods have been implemented for quantitative confocal phase imaging such as heterodyne \cite{RN226,RN223,RN202}, phase-locked \cite{RN227}, dual-phase \cite{RN228} and scanning-interferometer techniques \cite{RN146,RN206}, and optical surface profiling and refractive index mapping have been demonstrated \cite{RN224}. 
Synthetic optical holography (SOH) is a recently introduced modality for scanning quantitative phase imaging in scanning imaging systems \cite{schnell_synthetic_2014}, which uniquely combines technical simplicity and high imaging speed. In SOH, the scattered field from a local probe (focus) is superposed with a reference field with a phase which varies slowly across the acquired image. While scanning the focus across the sample, the detected light is recorded as a function of position, yielding an image exhibiting a fringe pattern much like that found in an off-axis Leith-Upatnieks hologram, that is, the system generates a synthetic hologram in a pointwise fashion. Amplitude and phase images can then be reconstructed by numerical methods developed for holography. Applied in a confocal system to record the coherently back-scattered signal, the phase images offer the transverse resolution of the confocal system while providing the high sensitivity to axial variations in surface profile with sub-nanometer depth resolution \cite{schnell_quantitative_2014,canales-benavides_accessible_2019,schnell_high-resolution_2020}. With phase and amplitude available, physics-based methods for numerical refocusing or aberration correction should be possible.  However, techniques from wide-field microscopy do not naively apply. As will be shown, the confocal geometry results in a rephasing factor of a different form to correct defocus.


Here we derive and demonstrate numerical refocusing of the coherently back-scattered scattered field in confocal quantitative phase imaging (confocal QPI). A forward model is derived for confocal microscopy in the backscattering geometry. A back-propagation operator is found via a pseudoinverse for the imaging operator.  An analytic solution is found that can be implemented as a rephasing in the Fourier domain. After developing the theory, results are demonstrated with a United States Air Force (USAF) target imaged with a confocal microscope with synthetic optical holography to obtain amplitude- and phase-resolved images. The numerically refocused data are compared to in-focus non-interferometric images.

\section{Confocal QPI}

Figure \ref{fig:holography_setup}a illustrates our confocal microscope setup using SOH. The expanded beam from a stabilized HeNe laser ($\lambda = \SI{632.8}{\nano\meter}$) is focused on the sample surface by microscope objective L1 (20x, 0.4 NA). The coherent back-scattered light $U_S(\vect{r})$ from the sample is collected by the microscope objective and focused onto the pinhole by lens L2 (f=100 mm). A pinhole diameter of $\SI{100}{\micro \meter}$ (approx. 5 times the size of the Airy disk) was chosen to suppress multiple scattering and stray light. At each position $\vect{r}=(x,y)$ of the sample, the scattered light  $U_S(\vect{r})$ is superposed at the detector with a reference field $U_R(\vect{r})=A_R e^{i\varphi_R(\vect{r})}$. The latter is generated by reflection of the illuminating beam at the reference mirror PZM. While the sample is rapidly scanned, the reference mirror is slowly translated at constant velocity, producing a slowly varying phase of the reference field $\varphi_R(\vect{r})$. By recording the detector signal $I(\vect{r})$ pixel-by-pixel, we obtain a synthetic hologram of the sample that is analogous to a plane-wave off-axis hologram (see ref. \cite{schnell_quantitative_2014} for details). The complex-valued scattered field, $U_S(\vect{r})$, is obtained using standard FT filtering. To allow for a direct comparison with non-interferometric (intensity) confocal images, we display the intensity of the complex-valued data $I_S(\vect{r}) = |U_S(\vect{r})|^2$. To produce in- and out-of-focus conditions, the sample can be moved in z by the piezo stage used for the sample scanning.

\section{Numerical Refocusing Method}

\begin{figure}
		\centering
		\includegraphics[width=0.95\linewidth, trim= 2.2in 6.15in 1.7in 1.65in, clip]{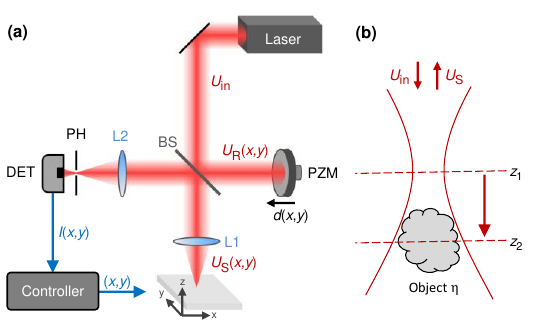}
		\caption[Implementation of synthetic optical holography in a confocal microscope for computational refocusing.]{Implementation of synthetic optical holography in a confocal microscope for computational refocusing. (a) Setup. (b) Confocal imaging of an object far from the focal plane.}	
		\label{fig:holography_setup}
	\end{figure}

Numerical refocusing can be achieved using the recovered complex signal and information about the nature of the confocal system. In confocal imaging of an object far from the focal plane, the source, aperture and lens produce a Gaussian beam propagating in the z-direction which comes to focus at position $z_1$ and illuminates the sample $\ket{\eta}$ located at position $z_2$ (Fig. \ref{fig:holography_setup}b). The elastically scattered field $\ket{S_1}$ from the object is collected by the same aperture as the incident field. The forward model is described by the imaging operator $P$, whose input is the object $\ket{\eta}$ and output is the scattered field $\ket{S_1}=P\ket{\eta}$. To refocus, we first reconstruct the object $\ket{\eta}$ from the scattered field $\ket{S_1}$ using the pseudoinverse of the imaging operator for the far-from-focus case, $P_1^+$. Then the forward operator for the near-focus case, $P_2$, is applied to the reconstructed object, resulting in an estimate of what the scattered field would have been if the focal plane was at position $z_2$ instead of $z_1$. Both operations are described by the refocusing operator $R_{12} = P_2P_{1}^+$, and the refocused scattered field $\ket{S_2}$ is then $\ket{S_2}=R_{12} \ket{S_1}$. In the following, we present how $R_{12}$ is derived.

The beam has peak amplitude in plane $z_1$, and its shape is $\tilde{g}(\vect{k}_\pll) = e^{-\alpha^2 \left| \vect{k}_\pll \right| ^2/2k^2}$ where $\vect{k}_\pll$ is a transverse Fourier domain variable, $k$ is the wavenumber, and $\alpha$ is a constant controlling the width of the beam \cite{ISAM}. The imaging operator may then be written as follows, where $k_z(\vect{k}_\pll)=\left(k^2- \left| \vect{k}_\pll 
\right| ^2\right)^{1/2}$ \cite{ISAM}:

	\begin{multline}
	\label{eq:Forward_Model} 
		P=\int dz\,d^2k'_\pll\, d^2k_\pll\, \ket{\vect{k}_\pll} U_o\tilde{g}(\vect{k}'_\pll)\tilde{g}(\vect{k}'_\pll-\vect{k}_\pll) \times \\ \frac{ie^{i(z-z_1)(k_z(\vect{k}'_\pll)+k_z(\vect{k}'_\pll-\vect{k}_\pll))}}{k_z(\vect{k}'_\pll)}\bra{\vect{k}_\pll}\bra{z}
	\end{multline}
 When $P$ operates on an object $\ket{\eta}$, the  transverse Fourier components $\vect{k}_\pll$ of the object are mapped onto the $\vect{k}_\pll$ components of the scattered field. Information about the object corresponding to evanescent waves  and any homogeneous waves outside the spatial bandpass of the system is lost upon propagation to the far field, creating a null space of the operator. Therefore, $P$ does not have a true inverse, so the pseudoinverse is calculated. However, the integral over $\vect{k}_\pll'$ makes this difficult to calculate. Instead, let us consider Eq. \ref{eq:Forward_Model} under a far field approximation. Note that aside from the constant $iU_0$, the integrand can be written as $f(\vect{k}_\pll,\vect{k}_\pll',z)e^{ig(\vect{k}_\pll,\vect{k}_\pll',z)}$, where $f(\vect{k}_\pll,\vect{k}_\pll',z)$ and $g(\vect{k}_\pll,\vect{k}_\pll',z)$ are real-valued. Since the $z-z_1$ term in the exponent is large compared to the Rayleigh range for values of $z$ that contribute to the integral, the phase oscillates quickly compared to the rest of the integrand as $\vect{k}_\pll'$ is varied. Therefore the method of stationary phase can be applied, with the stationary point being $\vect{k}_\pll'=\frac{\vect{k}_\pll}{2}$ \cite{erdelyi1955asymptotic}. Define $H_F\equiv U_o\pi\tilde{g}^2(\frac{\vect{k}_\pll}{2})$, and let $P_1$ be the imaging operator with the focal plane $z_1$. $P_1$ may then be approximated as

	\begin{equation}
	\label{eq:P1Asymptotic} 
		P_1\sim\int d^2k_\pll dz\ \ket{\vect{k}_\pll}\frac{H_F}{z-z_1}e^{i2k_z(\frac{\vect{k}_\pll}{2})(z-z_1)}\bra{\vect{k}_\pll}\bra{z}
	\end{equation}
Since $z$ is assumed to be far from the focal plane, $z-z_1$ is always nonzero. The pseudoinverse of Eq. \ref{eq:P1Asymptotic} yields the imaging operator for the far-from-focus case, 
	\begin{equation}
	\label{eq:P1+Asymptotic} 
		P^+_1\sim \int d^2 k_\pll\, dz\ \ket{\vect{k}_\pll}\ket{z}\frac{z-z_1}{H_F}e^{-i2k_z(\frac{\vect{k}_\pll}{2})(z-z_1)}\bra{\vect{k}_\pll}
	\end{equation}
Next the approximated imaging operator for the near-focus case, $P_2$, is calculated. In this case, the product of the Gaussians in Eq. \ref{eq:Forward_Model} is very sharply peaked around $\vect{k}_\pll'=\vect{k}_\pll /2$, so only values of $\vect{k}_\pll'$ close to $\vect{k}_\pll /2$ contribute to the integral. The non-Gaussian portion of the integrand from Eq. \ref{eq:Forward_Model} is therefore expanded around the point $\vect{k}'_\pll=\frac{\vect{k}_\pll}{2}$. Because this portion of the integrand is even with respect to $\vect{k}_\pll'$, all odd order terms are zero, including the first order term. Terms of higher order than one are neglected, leaving only the zeroth order term. Define $H_N\equiv U_o\pi\tilde{g}^2(\frac{\vect{k}_\pll}{2})\frac{k^2}{\alpha^2} = H_F \frac{k^2}{\alpha^2}$ and let $P_2$ be the imaging operator with focal plane $z_2$. $P_2$ may then be approximated as:
	\begin{equation}
	\label{eq:P2_Asymptotic} 
		P_2 \sim \int d^2k_\pll dz \ket{\vect{k}_\pll}\frac{H_N}{k_z\left(\frac{\vect{k}_\pll}{2}\right)}ie^{i2k_z(\frac{\vect{k}_\pll}{2})(z-z_2)}\bra{\vect{k}_\pll}\bra{z}
	\end{equation}
 The refocusing operator $R_{12}$ from plane $z_1$ to $z_2$ is:
	\begin{equation}
	\label{eq:refocus_general} 
		\ket{S_2}=R_{12}\ket{S_1}=P_{2}P_{1}^+\ket{S_1}	
	\end{equation}
 Substituting Eqs. \ref{eq:P1+Asymptotic} and \ref{eq:P2_Asymptotic} into Eq. \ref{eq:refocus_general} and simplifying yields:
	\begin{equation}
	\label{eq:refocus_simplified} 
		R_{12} = \int d^2k_\pll dz \ket{\vect{k}_\pll}\frac{k^2}{\alpha^2}\frac{z-z_1}{k_z(\frac{\vect{k}_\pll}{2})}ie^{i2k_z\left(\frac{\vect{k}_\pll}{2}\right)(z_1-z_2)}\bra{\vect{k}_\pll}
	\end{equation}
 Using Eq. \ref{eq:refocus_simplified}, the relationship between the out of focus field and the refocused field can be written in functional notation as:
\begin{multline}
	\label{eq:refocus_projected}
		S_2(\vect{r}_{\pll2}) = \frac{k^2}{\alpha^2} \int d^2k_\pll dz \left[ \int dr_{\pll1} \frac{1}{2\pi} e^{-i \vect{k}_{\pll} \cdot \vect{r}_{\pll1}} S_1(\vect{r}_{\pll1}) \right] \times\\  \frac{z-z_1}{k_z(\frac{\vect{k}_\pll}{2})}  ie^{i2k_z\left(\frac{\vect{k}_\pll}{2}\right)(z_1-z_2)} \frac{1}{2\pi}e^{i \vect{k}_{\pll}\cdot \vect{r}_{\pll2}}
	\end{multline} 
By studying Eq. \ref{eq:refocus_projected} we can see that broadly, the refocusing operation is accomplished by taking the Fourier transform of the out of focus field, multiplying by a rephasing term, and taking the inverse Fourier transform. As the assumption that the object was far from the initial focus was built into the problem, the integral over $z$ in Eq. \ref{eq:refocus_projected} may be limited to some region around $z_2$ or removed entirely; there is no explicit dependence on $z$ outside of the amplitude, which may be scaled arbitrarily. The weighting includes a factor of $\frac{1}{k_z(\frac{\vect{k}_\pll}{2})}$, which amplifies high frequency components while subduing lower frequencies. Some amplification is introduced according to the distance from the initial focal plane, as the original field decayed as $\frac{1}{z}$ as it propagated to the sample. In practice, however, the amplitude factors $
\frac{k^2(z-z_1)}{\alpha^2}$ can be neglected along with the integral over $z$, which is what is done in the following section. This result is in agreement with other known results for the wide-field case, where refocusing amounts to rephasing in the Fourier domain \cite{MmWave_Imaging,3D_Reconstruction}. However, some results express the new phase factor as quadratic in the $\vect{k}_\pll$ component \cite{Terahertz_Holography}. By performing an expansion of Eq. \ref{eq:refocus_simplified} in $\vect{k}_\pll$ and noting that the first order term contributes nothing, it can be shown that this quadratic rephasing in wide field is simply the second order expansion of the result derived here.


\section{Results}

\begin{figure}
		\centering
		\includegraphics[width=0.65\linewidth, trim= 0.8in 0.8in 4.0in 0.5in, clip]{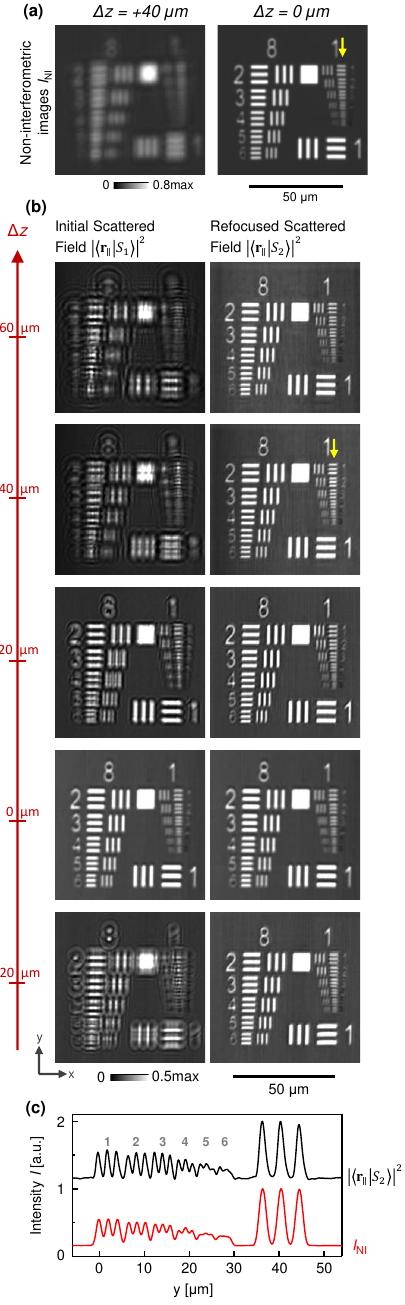}
		\caption[Demonstration of numerical refocusing with a reflective 1951-USAF test sample.]{(a) Non-interferometric images of the samples acquired out-of-focus and in-focus with the reference arm blocked. (b) Intensity images of the initial scattered field $\bra {\vect{r}_\pll}\ket{S_1}$ (left panel) and refocused scattered field $\bra {\vect{r}_\pll}\ket{S_2}$ (right panel) as obtained for different distances, $\Delta z$, of the sample from the focal plane. (c) Line profiles across the elements of group 9 (as indicated by arrows) for the refocused field intensity, $|\bra {\vect{r}_\pll}\ket{S_2}|^2$, $\Delta z = \SI{40}{\micro\meter}$ in (b) and the in-focus non-interferometric image, $I_\text{NF}$, in (a). Note: Due to imperfect alignment of the confocal system, the sample may have been slightly out of focus, yielding a slightly reduced resolution in the non-interferometric data.}	
        \label{fig:demo_refoc}
	\end{figure}

In Fig.~\ref{fig:demo_refoc} we demonstrate and verify our refocusing method with a reflective 1951 USAF resolution test sample that was positioned at varying distances $\Delta z = z_2 - z_1$ from the microscope’s focal plane. The test sample was fabricated by focused-ion beam milling of a 40 nm thick Au film located on a $\text{CaF}_2$ substrate, yielding highly reflective bars and numbers on a weakly reflective substrate. Note that group 9 is incorrectly labeled by ‘1’ due to an error in fabrication. The implementation of the refocusing algorithm was done in Matlab (Supplement 1, Note 1, code in ref. \cite{schnell_martin_computational_2023}). We first reconstruct the SOH hologram as described in previous work\cite{schnell_quantitative_2014}. We then apply a tapered-cosine window function to the image to pull the data near the image border to a single, mean value in a smooth transition, thus avoiding diffraction artifacts in the refocusing algorithm. We then perform a two-dimensional FT on the data, apply the refocusing operator $R_{12}$ from Eq. \ref{eq:refocus_simplified}, subsequently perform an inverse FT, and crop the previously windowed part of our data.

To validate our numerical refocusing method, we acquired the initial scattered field $\ip{\vect{r}_\pll}{S_1}$ for different out-of-focus depths $\Delta z$ ranging from $\SI{-20}{\micro\meter}$ to $\SI{60}{\micro\meter}$ (Fig.~\ref{fig:demo_refoc}b). In the initial scattered field interferometric images, the test pattern can already be recognized in bright color on a dark background (weakly reflecting substrate), together with light and dark rings characteristic of defocus in coherent imaging. Refocusing of the data consistently resolved all elements of group 8 and elements 1 to 4 of group 9 of the test pattern. This is consistent with the level of detail in the in-focus non-interferometric image (Fig.~\ref{fig:demo_refoc}a, $\Delta z=0$), confirming qualitatively the correct refocusing. In contrast, the out-of-focus non-interferometric image (Fig.~\ref{fig:demo_refoc}a, $\Delta z = \SI{40}{\micro\meter}$) does not resolve any but the largest group of bars of the pattern. This image cannot be numerically refocused because only the intensity information of the scattered field $|\ip{\vect{r}_\pll}{S_1}|^2$ was recorded.  This example demonstrates impressively the advantage of quantitative amplitude- and phase-resolved confocal imaging as provided by SOH, which allows for refocusing of out-of-focus data after image acquisition and well beyond the depth of focus of the microscope object ($\text{DOF}=\lambda/\text{NA}^2 \approx \SI{4}{\micro\meter}$ in our case). Note that refocusing also brings the phase information back into focus, yielding a sharp image of the optical topography of our test pattern, as it is illustrated in Supplement 1, Note 2.

In  Fig.~\ref{fig:demo_refoc}c we analyze the spatial resolution of the refocused data in more detail by extracting line profiles from the images in Fig.~\ref{fig:demo_refoc}a and \ref{fig:demo_refoc}b along the group 9 of the test pattern (as marked by the arrows). Utilizing the criterion of a $26\%$ drop between adjacent peaks (Rayleigh criterion), we find that the refocused scattered field $\ip{\vect{r}_\pll}{S_2}$ resolves all elements up to number 5 of group 9, which is equivalent to a spatial resolution of at least 812 line pairs/mm and near the theoretical diffraction limit of $\SI{1465}{lp/\milli\meter}$ (as given by $0.61\lambda/(\sqrt{2}\text{NA})$). In comparison to the in-focus recorded, non-interferometric data ,$I_\text{NI}$, we notice that refocusing has fully restored the spatial resolution of our instrument. 


\section{Discussion}

The numerically refocused image has the same bandwidth as the input data. In principle, numerical refocusing restores the spatial resolution up to the diffraction limited as determined by the bandwidth of the image passed by the optical system. In practice, the following constraints have to be considered. First, the signal-to-noise ratio of the detector signal limits the useful range of out-of-focus distances $\Delta z$ because (i) the scattered light falls off with $1/\Delta z$ due to the confocal depth discrimination ($1/\Delta z$ rather than $1/\Delta z^2$ because the field amplitude is detected in SOH) and (ii) the information of the scattered field is spread out over more pixels in the image for larger $\Delta z$, which increases noise. In Supplement 1, Note 3, we show the depth response curve of confocal SOH, validate that the amplitude signal in confocal SOH follows an inverse-linear scaling law with the out-of-focus distance, $\Delta z$, and determine that the greatest distance from which an object can be refocused is about $\SI{300}{\micro \meter}$ from the focal plane. Furthermore, we expect a vignetting effect for extreme out-of-focus distances where the physical size of the objective lens (aperture) acts as a spatial bandpass filter in the collection of the scattered light. We note that the aberration of the objective lens could be introduced as well, which could allow for correction of this aberration in confocal QPI.

\section{Conclusions}
We have demonstrated the numerical refocusing in monochromatic confocal microscopy where the elastically scattered light is detected. The method is based on the computation of the pseudoinverse of the imaging operator of the confocal imaging system and a subsequent re-imaging of the object to the desired focal plane. Numerical refocusing of phase-resolved confocal data opens the way for focused views on different planes of a specimen as well as for correction of out-of-focus data after image acquisition without the need to re-image. The method may be extended for tomographic imaging where it would be necessary to introduce a third degree of freedom for a full three-dimensional reconstruction, such as for example a rotation of the object. Numerical refocusing can also enable precise optical topography measurements over a wide range in depth, which could be useful for characterizing small details (e.g. surface defects) over a surface with large height variations.


\begin{backmatter}
\bmsection{Funding} Spanish Ministry of Science and Innovation under the Maria de Maeztu Units of Excellence Program (CEX2020-001038-M/MCIN/AEI/10.13039/501100011033); Ministerio de Ciencia e Innovación (MCIN/AEI/10.13039/501100011033) (PID2020-115221GA-C44, PID2021-123949OB-I00); ERDF—A Way of Making Europe.  The NSF I/UCRC Center for Freeform Optics (IIP-1822049 and IIP-1822026).



\bmsection{Disclosures} RH: attocube systems AG (F,C,P,R). All other authors declare no conflict of interest.


\bmsection{Data availability} Data underlying the results presented in this paper are available in Ref. \cite{schnell_martin_computational_2023}.

\bmsection{Supplemental document}
See Supplement 1 for supporting content. 

\end{backmatter}

\bibliography{microscopy}

\bibliographyfullrefs{microscopy}


\ifthenelse{\equal{\journalref}{aop}}{%
\section*{Author Biographies}
\begingroup
\setlength\intextsep{0pt}
\begin{minipage}[t][6.3cm][t]{1.0\textwidth} 
  \begin{wrapfigure}{L}{0.25\textwidth}
    \includegraphics[width=0.25\textwidth]{john_smith.eps}
  \end{wrapfigure}
  \noindent
  {\bfseries John Smith} received his BSc (Mathematics) in 2000 from The University of Maryland. His research interests include lasers and optics.
\end{minipage}
\begin{minipage}{1.0\textwidth}
  \begin{wrapfigure}{L}{0.25\textwidth}
    \includegraphics[width=0.25\textwidth]{alice_smith.eps}
  \end{wrapfigure}
  \noindent
  {\bfseries Alice Smith} also received her BSc (Mathematics) in 2000 from The University of Maryland. Her research interests also include lasers and optics.
\end{minipage}
\endgroup
}{}

\end{document}